\documentclass{elsart}

\usepackage{graphicx}

\usepackage{amssymb}

\begin{document}

\begin{frontmatter}



\title{Characterising a solid state qubit via environmental noise}

\author[label1]{J.F.Ralph}
\author[label2]{T.D.Clark}
\author[label2]{M.J.Everitt}
\author[label2]{H.Prance}
\author[label2]{P.Stiffell}
\author[label2]{R.J.Prance}
\address[label1]{Department of Electrical Engineering and Electronics, The University of Liverpool,
Brownlow Hill, Liverpool, L69 3GJ, United Kingdom.}
\address[label2]{School of Engineering, The University of Sussex,
Falmer, Brighton, BN1 9QT, United Kingdom.}

\begin{abstract}
We propose a method for characterising the energy level structure of a solid-state
qubit by monitoring the noise level in its environment. We consider a model
persistent-current qubit in a lossy resevoir and demonstrate that the noise
in a classical bias field is a sensitive function of the applied field.
\end{abstract}

\end{frontmatter}

In this paper, we propose a novel technique to characterise the
energy level structure of a solid state qubit and its spontaneous emission rate
into a lossy reservoir. The technique relies on the backreaction of a
solid state qubit on its environment and the incoherent transfer of
energy from a high frequency mode to a low frequency mode due to the
stochastic transitions of the qubit between energy eigenstates.

We consider a coupled system consisting of a model qubit and several
classical degrees of freedom. The qubit is coupled to two main electromagnetic modes:
a low frequency mode (classical bias field) that is used to control
the operation of the qubit, and a higher frequency mode that
is used to pump the qubit from the ground state to an excited
state. In addition, the qubit is assumed to be coupled to a lossy
reservoir, which represents the cavity that contains the qubit and
control fields. The reservoir provides a mechanism to allow the qubit
to dissipate energy and to induce spontaneous decays from an excited
state into the ground state. The loss mechanism is modelled using a
quantum trajectory approach \cite{Car93,Wis96}, corresponding to an
unravelling of the Markovian Master equation for the qubit reduced
density operator. In this paper, we choose the {\it quantum jumps}
approach \cite{Car93,Wis96}, which is suitable for modelling spontanteous
emission processes and is computationally efficient \cite{Wis99}.
Physically, this unravelling corresponds to the (irreversible)
absorption of any
spontaneously emitted photon on a time scale that is significantly
faster than any of the time scales present in the quantum system.
However, other unravellings of the Master equation can also be used
to generate discrete stochastic jump-like behaviour and each process
generates the usual reduced density evolution when averaged over an
ensemble \cite{Car93,Wis96,Wis99,Gis93}.

Although these unravellings
reproduce the Master equation evolution when averaged, they also
contain other behaviour at the individual system level, such as
chaotic-like behaviour that gives an indication of a classical limit
that is not present in the reduced density operator \cite{Spi94}.
Unfortunately, the general characteristics of evolution (jumps,
chaos, etc.) are not specific to the unravelling and a large class of
unravellings can have similar properties. As a result, the
individual trajectories are normally considered to be subjective
\cite{Wis96} by virtue of the fact that the predictive power of a
simulation is contained in the average evolution, but care is required
as to where the average is taken. In this paper, we consider the time-averaged
evolution of a classical oscillator that is coupled reactively to a
qubit that is described by a quantum jump model, as described in
\cite{Wis99}. We demonstrate that the time-averaged behaviour of the
classical oscillator can be used to characterise the energy level
structure of the qubit and the rate of spontaneous decay into the
reservoir. The behaviour is
dependent on the presence of stochastic jump-like behaviour in the
qubit, but it is not specific to the unravelling used.

The model qubit studied in this paper is a superconducting persistent
current qubit proposed by Orlando et al. \cite{Orl99}. It is of
particular interest because the states that would be used in an
operational qubit device correspond to macroscopic current states, of
the type that have been used in a series of experiments to demonstrate
quantum properties of superconducting circuits: avoided-crossings in
the energy level structure \cite{Fri00} and coherent
oscillations of macroscopically distinct states \cite{Chi03}. Because of
the macroscopic nature of the current states, it is also a system
where the backreaction of the quantum device on the classical
environment can be significant \cite{Ral01}. Although a persistent
current qubit is studied in this paper,
the same type of techniques could be adapted for other solid state
qubits where the currents and/or voltages are comparable with those
that are present in the classical control fields. The
two-state Hamiltonian used in this paper is quite general and
capacitive coupling between charge quibts could replace the inductive
coupling used here. Small capacitance charge qubits can generate significant
voltages from a small number of charges. A single electron on a
charge qubit with a capacitance of $C\simeq 10^{-15}$ F would
produce a voltage of around $V\simeq 0.2$ mV, which is large enough
to produce a significant backreaction on a classical field.

The superconducting circuit described in reference \cite{Orl99} has
been proposed as a candidate for the implementation of quantum
processing in solid state devices. It is one of the most developed
forms of persistent current qubit and similar devices having been
used in recent experiments providing evidence for macroscopic coherent
oscillations \cite{Chi03}. In this paper, we use the description of
the system, together with the appropriate parameter values, given in
reference \cite{Orl99} (see Figure 1). In this description, the qubit
can be simplified to a two-state model corresponding to two current states
which differ by approximately 600 nA. The Hamiltonian for this system
is given by,
\begin{equation}
\hat{H}_{qu}(\Phi_{x1},\Phi_{x2})=\left(\begin{array}{cc}
F(\Phi_{x1},\Phi_{x2})& -B(\Phi_{x1},\Phi_{x2}) \\
-B(\Phi_{x1},\Phi_{x2})& -F(\Phi_{x1},\Phi_{x2})
\end{array}\right)
\end{equation}
where
\begin{equation}
F(\Phi_{x1},\Phi_{x2})=r_1\left(\frac{\Phi_{x1}}{\Phi_0}\right)+r_2\left(\frac{\Phi_{x2}}{\Phi_0}\right)
\end{equation}
\begin{equation}
B(\Phi_{x1},\Phi_{x2})=\frac{t_1+s_1\left(\frac{\Phi_{x1}}{\Phi_0}\right)}{1-\eta
\sqrt{\frac{E_J}{E_C}}\left(\frac{\Phi_{x2}}{\Phi_0}\right)}
\end{equation}
and where the two principal control fields, the magnetic flux biases
$\Phi_{x1}$ and $\Phi_{x2}$, are used in the operation of the qubit and
may be time-dependant, and $\Phi_0=h/2e=2\times 10^{-15}$ Wb.
The circuit specific constants are taken from
\cite{Orl99}: $r_1 = 2\pi E_J\sqrt{1-\frac{1}{4\beta^2}}$,
$r_2=r_1/2$, $s_1=0$, $t_1=0.001 E_J$, $\eta=3.5$, $\beta=0.8$,
$E_J\equiv 200$ GHz, $E_C=E_J/80$. For the purposes of this paper, we
will set $\Phi_{x2}=0$ for convenience and introduce a relatively weak
time-dependent field in $\Phi_{x1}$ to induce transitions between the
ground state and the excited state. We will typically use a
time-dependent field of the form $\Phi_{hf1}(t)= 0.00003\Phi_0
\cos(\omega t)$, with a frequency of 500 MHz. This applied field is
the high frequency mode that we will use to drive the qubit, which - in turn -
dissipates into the environment and transfers energy to the low
frequency oscillator mode via an inductive backreaction. In theory,
this drive could be coupled into the qubit via a circuit similar to
the bias oscillator shown in Figure 1, but in practice it is more
usual to drive the qubit devices at microwave frequencies using a
coaxial waveguide \cite{Fri00}. For this reason, the high frequency mode is not
modelled explicitly in this paper. The drive frequency is above the
minimum splitting of the two states (which is 400 MHz) and excites transitions
at a static flux bias of around $\Phi_{x1}=0.00015\Phi_0$, as shown in
the insert in Figure 1. The insert in Figure 1 also shows a weak
two photon transition, even for the very small amplitude drive used in
this paper. Larger drive amplitudes can give rise to even higher order
transitions due to the large nonlinearities present in Josephson
junction circuits \cite{Cla98}.
\begin{figure}
\begin{center}
\includegraphics[height=8cm]{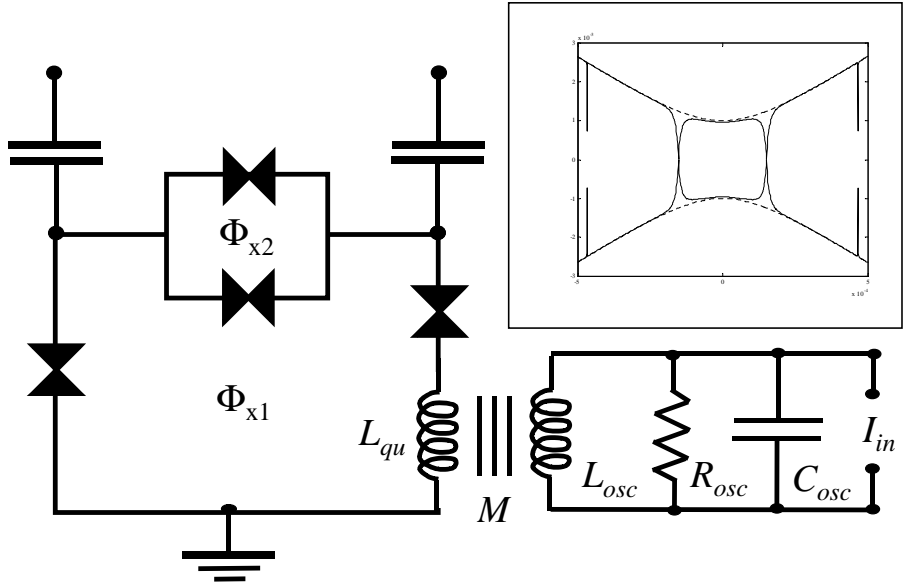}
\end{center}
\caption{Schematic diagram of persistent current qubit \cite{Orl99}
inductively coupled to a (low frequency) classical oscillator. The
insert graph shows the time-averaged (Floquet) energies as a function of
the external bias field $\Phi_{x1}$ for the parameters
given in the text.}
\end{figure}

The qubit circuit is designed so that the inductance of the superconducting
loop is negligible when compared with the effective inductance
generated by the series Josephson junctions in the loop
($L_J=\Phi_0/2\pi I_0$, where $I_0$ is the junction critical
current). This means that the energy levels and the persistent current
states of the circuit are dominated by the Josephson junctions rather
than the geometrical inductance of the ring, $L_{qu}\simeq 10$ pH
\cite{Orl99}. However, the inductance
does play an important role when determining the coupling between the
ring and the the external fields. In inductively coupled circuits, the
magnetic fluxes are related to the currents flowing through the
inductors by the inductances and the mutual inductances. For the qubit
coupled to the low frequency oscillator, this gives
\begin{equation}
\Phi_{qu}=L_{qu} I_{qu}+M I_{osc}
\end{equation}
\begin{equation}
\Phi_{osc}=L_{osc} I_{osc}+M I_{qu}
\end{equation}
where $M$ is the mutual inductance, $\Phi_{osc}$ is the magnetic flux in the oscillator
and the oscillator is characterized by a capacitance $C_{osc}$, and an
inductance $L_{osc}$. In the absence of dissipation (represented by
the parallel resistance in the oscillator circuit) the effective classical Hamiltonian for the
combined system can be written in the form \cite{Ral92},
\begin{equation}
H=\frac{Q_{osc}^2}{2C_{osc}}+\frac{\Phi_{osc}^2}{2L_{osc}}-\Phi_{osc}I_{in}
+\left<\hat{H}_{qu}(\mu\Phi_{osc}+\Phi_{hf1}(t),\Phi_{x2})\right>
\end{equation}
where $I_{in}$ is the external current applied to the oscillator
and the coupling coefficients are given by $K^2=M^2/L_{qu}L_{osc}$
and $\mu=M/L_{osc}$. An additional effect of the inductive coupling is
to shift the effective geometric inductance of the superconducting loop by a
factor $(1-K^2)$, $L_{qu}\rightarrow L_{qu}(1-K^2)$ \cite{Ral92}.

In the presence of dissipation, the classical equation of motion for
the oscillator becomes
\begin{equation}\label{osc}
C_{osc} \frac{d^2\Phi_{osc}}{dt^2}+\frac{1}{R_{osc}} \frac{d\Phi_{osc}}{dt}+\frac{\Phi_{osc}}{L_{osc}}=
I_{in}+\frac{\mu\left<\hat{I}_{qu}(\mu\Phi_{osc}+\Phi_{hf1}(t),\Phi_{x2})\right>}{(1-K^2)}
\end{equation}
where we have inserted the parallel resistance $R_{osc}$, and the qubit
screening current in the SQUID ring is calculated at each point in
time from the expectation value of the qubit screening current
operator $\hat{I}_{qu}$ over the instantaneous wavefunction (i.e. a pure state)
of the qubit using the bare (unrenormalised) inductance of the
qubit, $L_{qu}$. The use of the expectation value in a classical
equation of motion is a reasonable approximation in situations where the
oscillator has a lower frequency than any of the frequencies present
in the qubit and where any quantum fluctuations present in the
oscillator are negligible compared to its classical evolution.
It is consistent with the general approach to
quantum trajectories and the classical limit,
and retains the subjective nature of the trajectories because the
exact details of the quantum evolution of the qubit cannot be
reconstructed from the time-averaged classical behaviour of the
oscillator. Here, we use a noise-driven oscillator with a resonant
frequency of $\omega_{osc} = 1/\sqrt{L_{osc}C_{osc}}= 300$ MHz, a capacitance of
$C_{osc}=1$ nF, and a quality factor of
$Q_{osc}=\omega_{osc}R_{osc}C_{osc}=188$. The oscillator frequency is
chosen to be 300 MHz so that there are no harmonic relationships
between the oscillator and any of the frequencies within the region of
interest. The input current that is used to drive the oscillator
$I_{in}=I_{dc}+I_{noise}(t)$ consists of a dc current (which can be used to
bias the qubit, $\Phi_{dc}=L_{osc}I_{dc})$ and a time-varying component
due to Johnson noise in the resistor at finite temperature, $T=4$ Kelvin.
(The noise need not be thermal, but it is a useful generic model for
experimental noise because electronic noise is often characterised in
terms of an effective noise `temperature'). The
coupling between the qubit and the oscillator is very weak, $K =
0.01$ giving $\mu = 0.002$, to prevent any distortion of the apparent
qubit resonance due to a static backreaction \cite{Ral01}.

The quantum evolution of the qubit is determined by three competing
effects: the high frequency drive field, the noise-driven fluctuations in the coupled
oscillator, and the effect of spontaneous emission of the qubit into
the lossy reservoir. The drive field will induce transitions (Rabi oscillations) in the qubit when
the bias field is close to a quantum resonance. The noise present
in the low frequency oscillator will introduce dephasing (destroying
the absolute phase reference of any coherent oscillations) and a small
amount of dissipation due to the presence of the resistive element in
the oscillator circuit (the second effect is small by virute of the
weak coupling). The spontaneous emission process will introduce
quantum jumps that project into the instantaneous ground state
of the qubit. During time intervals where no spontaneous decay is
produced, the evolution contains terms due to coherent Schr\"{o}dinger
evolution and non-Unitary terms, as described in reference
\cite{Wis99}. During a small - but finite - time interval, $dt$, the
probability of a spontaneous decay occuring is
$\gamma\left<\hat{c}^{\dagger}\hat{c}\right>dt$, where $\hat{c}^{\dagger}$ and $\hat{c}$ are
the raising and lowering operators for the qubit states
respectively. If a spontaneous decay occurs, an operator
$\hat{\Omega}_{1}(dt)=\sqrt{\gamma dt}\:\hat{c}$ is applied to the qubit
wavefunction. If no spontaneous decay occurs, an evolution operator
$$\hat{\Omega}_{0}(dt)=1-\frac{i}{\hbar}\hat{H}_{qu}dt-\frac{\gamma}{2}\hat{c}^{\dagger}\hat{c} dt$$
is applied. The decay rate $\gamma$ is fixed by the coupling of
the qubit to the lossy reservoir.
\begin{figure}
\begin{center}
\includegraphics[height=10cm]{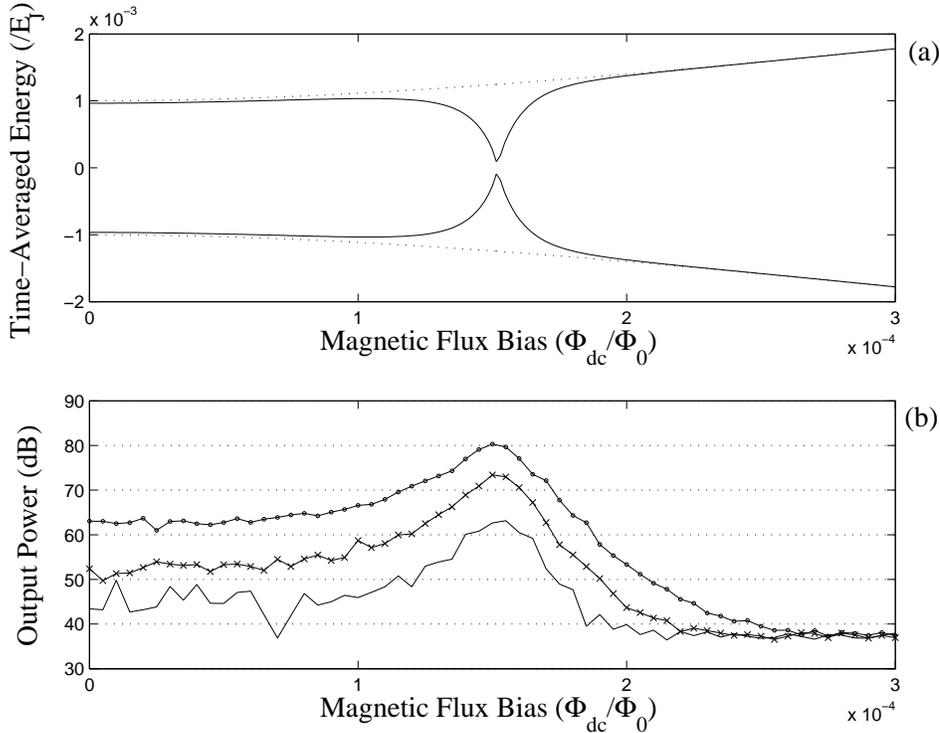}
\end{center}
\caption{(a) Close-up of the time-averaged (Floquet) energies of the
single photon resonance (500 MHz) - solid lines - with the
time-independent energies given dotted lines. (b) The output power of
the low frequency oscillator at 300 MHz, as a function of the static
magnetic flux
bias: $\gamma = 0.005$ per cycle (solid line), $\gamma = 0.05$ per
cycle (crosses), $\gamma = 0.5$ per cycle (circles). The other parameters are
given in the text}
\end{figure}

The evolution of the system is determined by numerical integration of
the equations of motion (classical and quantum jump), using random
initial states and ignoring the initial transient motion. The output
is chosen to be the voltage fluctuations in the classical oscillator,
which can either be characterised by the time-averaged root mean
square (RMS) power, or the power spectral density near a particular
frequency. Figure 2
shows three examples of the peak power in the main oscillator
resonance (near to 300 MHz) for different values of the spontaneous
decay rate $\gamma= 0.005, 0.05, 0.5$ per oscillator cycle. In each of
the graphs there is an asymmetry either side of the resonance. This is
due to the fact that the resonance is quite near to the minimum
splitting point and the energy eigenstates contain a significant
contribution from both current states near the minimum splitting (to the
left of the graph) and are predominately one current state on the
right of the resonance. Moving away from the minimum splitting point
(increasing the frequency of the applied field) restores the symmetry
of the peaks. (The apparent noise on the bottom graph is due to the
fact that the averaging process is limited in time and the damping
rate is sufficiently low that comparatively few quantum jumps are
seen).

In each of the cases shown in Figure 2, there is a significant
increase in output power near to the resonance, between 10 and 20 dB
between the left and the peak and between 20 and 40 dB between the
right and the peak. When the spontaneous decay rate is reduced the
power gain is reduced. Interestingly, there is no gain in output
power at the resonant frequency of the oscillator when there
are no spontaneous emissions ($\gamma = 0$) and the output power is not a
function of the applied static flux. This is because the discontinuous
jumps in the qubit are coupled back to the oscillator, and act as an
additional source of noise, thereby increasing the output power.
Near to the resonance, when the qubit is
more likely to be in the excited state, and therefore more likely to
undergo a quantum jump, the noise introduced by the jumps
is enhanced. Clearly, the output voltages
near the peaks shown in Figure 2 are significantly above the thermal
fluctuations in the oscillator (4 Kelvin $\simeq$ 40 dB in the units used here)
and their detection would allow the magnetic flux bias of
the resonance to be determined. The model does not include any direct
coupling between the low frequency and the high frequency modes, which
is likely to be present in an experimental system, but this coupling
will be independant of the static flux bias, so it is only the changes
in the output power as a function of the static flux that concern us here.

By detecting the position of the peak in the output voltage/power for different
frequencies of the drive field, it should be possible to
determine the energy level separations of the two states as a function
of static magnetic flux. The accuracy of the flux position would
be determined by the width of the resonance, and consequently the
amplitude of the applied drive field. The larger the high frequency
drive, the broader the resonance. For the example studied here the
flux accuracy is of the order of $10^{-5}\Phi_{0}$. The accuracy of
the frequency applied to the qubit is determined by the stability of the
classical (mono-frequency) source, but at the frequencies used a typical
value would be of the order of a few Hz. However, it should be noted that the
conventional method for applying high frequency fields (coaxial
waveguides \cite{Fri00}) is not best suited to this process because
the coupling between the qubit device and the applied drive field
may vary with frequency. This will not change the position of the
output peak, but it will change the shape of the peak because the
strength of the coupling will alter the width of the resonance.
\begin{figure}
\begin{center}
\includegraphics[height=18cm]{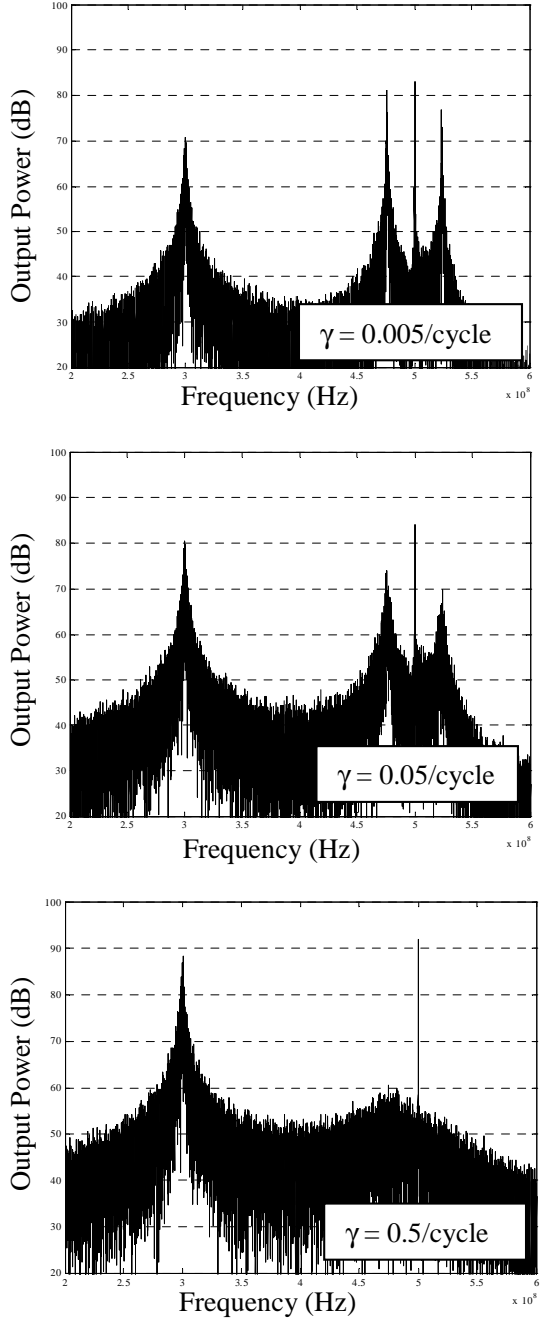}
\end{center}
\caption{Power spectral density for the low
frequency oscillator at the resonance point
($\Phi_{dc}=0.00015\Phi_0$) for the three spontaneous decay rates
shown in Figure 2: $\gamma = 0.005, 0.05, 0.5$ per cycle. The other parameters are
given in the text}
\end{figure}

Figure 3 shows the power spectral density of the output from the
oscillator for the three different decay rates shown in Figure 2.
The main resonance can be seen in each graph at 300 MHz, as well as
the fraction of the 500 MHz drive that couples directly from the high
frequency mode to the low frequency mode via the qubit. The other
peaks that can be seen either side of the 500 MHz peak in the first
two graphs are sidebands generated by mixing the Rabi oscillation
frequency with the drive frequency. The Rabi oscillation frequency
for the case considered here is much less than the resonant frequencies
of the modes, $f_{Rabi}\simeq 25$ MHz. In the first two graphs, the
coupling to the lossy reservoir is sufficiently low for the Rabi
oscillations to occur (more or less) undisturbed by the quantum
jumps. The larger linewidth in the second graph indicating the larger
spontaneous decay rate. In the third graph, there is no evidence of these Rabi
sidebands because the average spontaneous emission frequency
is much smaller than the Rabi frequency. Whilst the large signal gain
shown in Figure 2, is an indication of the presence of a quantum
transition, it does not indicate that the behaviour of the qubit is
dominated by coherent Schr\"{o}dinger evolution. The presence of
Rabi sidebands is a good indication that the
behaviour of the system is predominantly coherent and the linewidth of these sidebands
provides an indication of the damping rate. The position of
the sidebands relative to the drive frequency also gives a good estimate of
the Rabi frequency.

In this paper, we have discussed the effect of resonant transitions in
a model qubit on a low frequency oscillator. The qubit is coupled to
two modes and a lossy reservoir: one high frequency mode used to drive
the transitions between energy eigenstates of the qubit, and one low
frequency oscillator that is used to control the bias field for the
qubit. The coupling to the lossy reservoir induces spontaneous decay of
the qubit from an excited state to its instantaneous ground
state. Near a quantum mechanical resonance between the high frequency
mode and the qubit, the qubit is more likely to be in an excited state
and quantum jumps introduced by the spontaneous emission are more
frequent. The discontinuous jumps in the qubit are coupled to the
low frequency oscillator via the mutual inductance and give rise to
significant increases in the output power detected in the oscillator
when the qubit is biased near to a quantum resonace. We have
investigated the effiect of this backreaction on the low frequency
oscillator and suggested how this power gain could be used to
characterise the energy level structure of the qubit and its coupling
to the lossy reservoir.

The authors would like to thank the Engineering and Physical Science
Research Council (EPSRC) Quantum Circuits Network for their generous support.



\begin{thebibliography}{00}
\bibitem{Car93}
H.J.Carmichael, `An Open System Approach to Quantum Optics' (Lecture
Notes in Physics, Vol.18), Springer-Verlag, Berlin, 1993;
\bibitem{Wis96}
H.M.Wiseman, {\it Quant. Semiclass. Opt.} {\bf 8}, 205 (1996).
\bibitem{Wis99}
H.M.Wiseman, G.E.Toombes, {\it Phys. Rev. A} {\bf 60}, 2474 (1999).
\bibitem{Gis93}
N.Gisin, I.C.Percival, {\it J. Phys. A} {\bf 26}, 2233 (1993);
N.Gisin, I.C.Percival, {\it J. Phys. A} {\bf 26}, 2246 (1993);
G.C.Hegerfeldt, {\it Phys. Rev. A} {\bf 47}, 449 (1993).
\bibitem{Spi94}
T.P.Spiller, J.F.Ralph, {\it Phys. Lett. A} {\bf 194}, 235 (1994);
T.A.Brun, I.C.Percival, R.Schack, {\it J.Phys. A} {\bf 29} 2077
(1996).
\bibitem{Orl99}
T.P.Orlando, J.E.Mooji, L.Tian, C.H. van der Wal, L.S.Levitov,
S.Lloyd, J.J.Mazo, {\it Phys. Rev. B} {\bf 60}, 15398 (1999).
\bibitem{Fri00}
J.R.Friedman, V.Patel, W.Chen, S.K.Tolpygo, J.E.Lukens,
{\it Nature} {\bf 406}, 43 (2000);
C.H. van der Wal, A.C.J. ter Haar, F.K.Wilhem, R.N.Schouten,
C.J.P.M.Harmans, T.P.Orlando, S.Lloyd, J.E.Mooij, {\it Science}
{\bf 290}, 773 (2000).
\bibitem{Chi03}
I.Chiorescu, Y.Nakamura, C.J.P.M.Harmans, J.E.Mooij, {\it Science}
{\bf 299}, 1869 (2003).
\bibitem{Ral01}
J.F.Ralph, T.D.Clark, M.J.Everitt, P.Stiffell,
{\it Phys. Rev. B.} {\bf 64}, 180504 (2001));
J.F.Ralph, T.D.Clark, M.J.Everitt, P.Stiffell, R.J.Prance, H.Prance,
Proceedings of SPIE conference `Quantum Computing III', SPIE Vol. 4732
Eds. A.R.Pirich, E.W.Taylor, E.Donkor (2002).
\bibitem{Cla98}
T.D.Clark, J.Diggins, J.F.Ralph, M.J.Everritt, R.J.Prance, H.Prance, R.Whiteman,
{\it Ann. Phys.} {\bf 268}, 1 (1998); M.J.Everitt, T.D.Clark, P.Stiffell,
H.Prance, R.J.Prance, A.Vourdas, J.F.Ralph, {\it Phys. Rev. B} {\bf 64}, 184517 (2001).
\bibitem{Ral92}
J.F.Ralph, T.P.Spiller, T.D.Clark, R.J.Prance, H.Prance, {\it Int. J. Mod. Phys. B} {\bf 8}, 2637 (1994);
J.Diggins, J.F.Ralph, T.D.Clark, T.P.Spiller, R.J.Prance, H.Prance, {\it Phys. Rev. E} {\bf 49}, 1854 (1994);
T.D.Clark, J.F.Ralph, R.J.Prance, H.Prance, J.Diggins, R.Whiteman, {\it Phys. Rev. E} {\bf 57}, 4035 (1998).

\end{thebibliography}
\end{document}